\documentclass[fleqn,10pt]{SelfArx} 

\usepackage[english]{babel} 

\usepackage{lipsum} 
\usepackage{afterpage}

 \definecolor{color1}{RGB}{0,0,110} 
\definecolor{color2}{RGB}{0,20,80} 

\usepackage{hyperref} 
\hypersetup{hidelinks,colorlinks,breaklinks=true,urlcolor=color2,citecolor=color1,linkcolor=color1,bookmarksopen=false,pdftitle={Title},pdfauthor={Author}}

\JournalInfo{Uploaded to arXiv.org} 
\Archive{} 

\PaperTitle{Controlling ionic current through a nanopore by tuning pH: a Local Equilibrium Monte Carlo study} 

\Authors{D\'avid Fertig\textsuperscript{1}, M\'onika Valisk\'o\textsuperscript{1},  Dezs\H{o} Boda\textsuperscript{1,2}*} 
\affiliation{\textsuperscript{1}\textit{Department of Physical Chemistry, University of Pannonia,  P. O. Box 158, H-8201 Veszpr\'em, Hungary}} 
\affiliation{\textsuperscript{2}\textit{Institute of Advanced Studies K\H{o}szeg (iASK), Chernel u. 14, H-9730 K\H{o}szeg, Hungary}} %
\affiliation{*\textbf{Corresponding author}: dezsoboda@gmail.com} 

\newcommand{\keywordname}{Keywords} 

\Abstract{The purpose of this work is to create a model of a nanofluidic transistor which is able to mimic the effects of pH on nanopore conductance.  The pH of the electrolyte is an experimentally controllable parameter through which the charge pattern can be tuned: pH affects the ratio of the protonated/deprotonated forms of the functional groups anchored to the surface of the nanopore (for example, amino and carboxyl groups). Thus, the behavior of the bipolar transistor changes as it becomes ion selective in acidic/basic environments. We relate the surface charge to pH and perform particle simulations (Local Equilibrium Monte Carlo) with different nanopore geometries (cylindrical and double conical). The simulations form a self consistent system with the Nernst-Planck equation with which we compute ionic flux. We discuss the mechanism behind pH-control of ionic current: formation of depletion zones.}

\begin{document}

\flushbottom 

\maketitle 


\thispagestyle{empty} 



\section{Introduction}

Nanopores are nano-scale holes in synthetic membranes made of, for example, silicon, graphene, or plastic \cite{abgrall_book,iqbal_book_2011,zhang_csr_2018} that allow the flow of ions from one side of the membrane to the other.
These systems can be considered as nanodevices that provide an output signal (ionic current) as a response to voltage or ionic concentration difference.
They can be used for energy conversion \cite{hsu_jps_2017,gillespie_nl_2012} or as biosensors \cite{vlassiouk_jacs_2009,madai_pccp_2018}.
Devices that allow the control of current via tuning additional system parameters (electric field, electrolyte composition, geometry, etc.) are essential building blocks of nanofluidic circuitries \cite{abgrall_book}. 
We define this behavior as transistor-like behavior.

Control can be realized by tuning the electric field inside the nanopore because changing the electrostatic energy of ions determines the probability of residence of ions in the pore.
Of course, chemical and steric interactions can also be used for control, but our main interest is the electrostatic control here. 
Electric field can be tuned directly by setting electrical potential at embedded electrodes \cite{burgmayer_jacs_1982,Fan_PRL_2005,karnik_nl_2005,karnik_apl_2006,horiuchi_loac_2006,gracheva_acsnano_2008,kalman_am_2008,Kalman_BJ_2009,cheng_acsnano_2009,tybrandt_natcomm_2012,Lee_NS_2015,Fuest_nl_2015,Fuest_AC_2017}.
Electric field can also be changed by manipulating the surface charge pattern on the surface of the nanopore \cite{kuo_l_2001,stein_prl_2004,Siwy_2004,singh_jap_2011,jiang_pre_2011,singh_sab_2016,Singh_PCCP_2017}.

The charge pattern can be manipulated with chemical methods by anchoring functional groups to the pore wall \cite{siwy_prl_2002,Kim_2010,siwy_csr_2010,Ali_ACSnano_2012,gibb_chapter_2013,Duan_bmf_2013,guan_nanotech_2014,Fuest_AC_2017}.
Surface charge can also be modulated with pH if protonation/deprotonation of the functional groups is pH sensitive \cite{wang_rm_2009,xue_jpcc_2014,lin_ac_2015} such as pH-responsive polyethylene terephthalate (PET) and polyimide (PI) nanopores \cite{siwy_nim_2003,lepoitevin_acis_2017,Lin_2018}.
The goal of this paper is to provide a modeling study for understanding the mechanisms by which pH can influence the charge pattern, and, furthermore, charge patterns influence ionic transport.

The main motivation of our study is the work of Kalman et al.\ \cite{kalman_am_2008} who studied a biconical PET nanopore functionalized with carboxyl (n-region) and amino (p-region) groups in a symmetric way: ``pnp'' (see Fig.\ \ref{fig1}).
They showed pH-dependent experimental current-voltage curves and related their results to assumed pH-sensitive charge patterns using Poisson-Nernst-Planck (PNP) calculations. 
We used their symmetric charge pattern arrangement here and in our previous study \cite{madai_pccp_2018} in order to avoid asymmetric current-voltage curves.
This way, the sign of voltage does not enter our calculations as an additional parameter.
Furthermore, this symmetric charge pattern is usual in the nanopore literature \cite{daiguji_nl_2004b,gracheva_acsnano_2008,kalman_am_2008,nam_nl_2009,cheng_csr_2010,singh_sab_2016}, although asymmetric pores are more abundant due to their rectification properties.
Most of the experimental studies considering pH-regulated systems refer to rectifying nanopores.

When the pore wall is treated with chemicals that can carry both positive and/or negative charges depending on pH (zwitterionic \cite{yameen_jacscomm_2009}, amphoteric \cite{Ali_ACSNano_2009,perezmitta_ns_2016}), rectification properties can be tuned with pH. 
In the case of zwitterionic polymers, an inversion of rectification can be produced: the pore is open at different signs of the voltage at acidic and basic pH \cite{yameen_jacscomm_2009}.
In a nanopore, where polyprotic acidic groups are attached to the wall of the pore, the endgroups have three states as a function of pH: neutral, partially charged, and fully charged. 
This corresponds to three distinct rectification levels \cite{Ali_LangLett_2009,yameen_chemcomm_2010}.
In a glass nanopipette pore, pH can tune not only the protonation properties of the attached amino groups, but also silanol groups of the glass allowing diverse rectification behavior \cite{liu_analchem_2012}. 

An hourglass shape PET nanopore whose one side is grafted with poliacrylic acid becomes charged at large pH resulting in a unipolar rectifying nanopore. 
Also, amino acids can change their conformations between coiled and stretched states thus producing a switch behavior \cite{hou_advmat_2010}.
A DNA-modified nanopore also allows switch behavior by DNA strands meshing when they contain both positive and negative groups at appropriate pH thus closing the pore \cite{buchsbaum_jacs_2014}.

Modeling studies of pH-control require knowledge of the surface charge on the nanopore's wall.
One strategy is using PNP to solve the reverse problem: on the basis of experimental current-voltage curves PNP provides estimates for the surface charges \cite{Ali_ACSNano_2009,ali_acsami_2015,ali_sab_2017,Ali_AC_2018}.
The other strategy, which we use here, is to relate pH to protonation/deprotonation degrees, and, thus, surface charges, through dissociation equilibrium \cite{kalman_am_2008,cruzchu_2009}.

There are several studies using this approach.
Yeh et al.\ \cite{yeh_analchem_2013} used PNP and Navier-Stokes (NS) to study pH-dependent ion transport through cylindrical glass nanopores.
In the study of Hsu et al.\ \cite{hsu_analchem_2017}, a pH-tunable zwitterionic surface influenced rectification in a bullet-shape nanopore.
The PNP-NS study of Lin et al.\ \cite{lin_ac_2015} also considered pH-dependent rectification of conical nanopores functionalized with polyelectrolyte brushes.
A model of a pH-regulated field effect transistor describes the effect of pressure on streaming conductance and zeta-potential values at different voltages and pH \cite{xue_jpcc_2014}.

Most of the papers cited so far refer to asymmetric, rectifying nanopores.
The symmetric arrangement used in this paper was also considered in modeling studies.
In a series of papers, Gracheva et al.\ \cite{gracheva_acsnano_2008,Gracheva_JCE_2008,nikolaev_jce_2014} studied a nanopore through a semiconductor membrane, while Park et al.\ \cite{park_mfnf_2015} studied a model called double-well nanofluidic channel.
These authors considered a parameter range similar to ours (large surface charges and narrow pores) and the qualitative behavior reported by them is similar to the model studied by us here and in our previous study \cite{madai_pccp_2018}.

The mechanism of ion exclusion (formation of depletion zone) depends on the behavior of the double layers formed at the pore wall in the radial dimension.
If these double layers extend into the pore and prevent the formation of a bulk electrolyte in the centerline of the pore, depletion zones can be produced even for short pores.
The depth of the depletion zone depends on the strength of the surface charge as discussed here and in our previous study \cite{madai_pccp_2018}. 
This happens when the width of the double layer (Debye-length, $\lambda_{\mathrm{D}}$) is measurable to pore radius ($R_{\mathrm{pore}}$).
Here, we use narrow pores with relatively large concentrations and surface charges.
A detailed discussion on the effect of these parameters can be found in Ref.\ \cite{madai_pccp_2018}.

Other studies considered wider and longer pores at lower concentrations keeping the $R_{\mathrm{pore}}/\lambda_{\mathrm{D}}$ ratio close to 1 \cite{daiguji_nl_2005,singh_jap_2011b,singh_lc_2012,Singh_PCCP_2017}.
While the different ways of creating depletion zones in those cases and a scaling behavior as a function of $R_{\mathrm{pore}}/\lambda_{\mathrm{D}}$ were discussed in our previous work \cite{madai_pccp_2018} in detail, the effect of pH was briefly mentioned in that work.
Here, we provide a systematic analysis of pH-control of ionic transport through a model nanopore-based transistor.

\section{Model and method}

\subsection{Nanopore model}

The system  is composed of two baths separated by a membrane that is penetrated by a pore.
The system has a rotational symmetry around the axis of the pore ($z$-axis), therefore, the solution domain is a cylinder of 30 nm width and 18 nm radius.
The dimensions of the computational domain were large enough to provide current data independent of system size at the given concentration (0.1 M), but small enough to make fast calculations possible.

\begin{figure}[t]
\begin{center}
\rotatebox{0}{\scalebox{0.45}{\includegraphics*{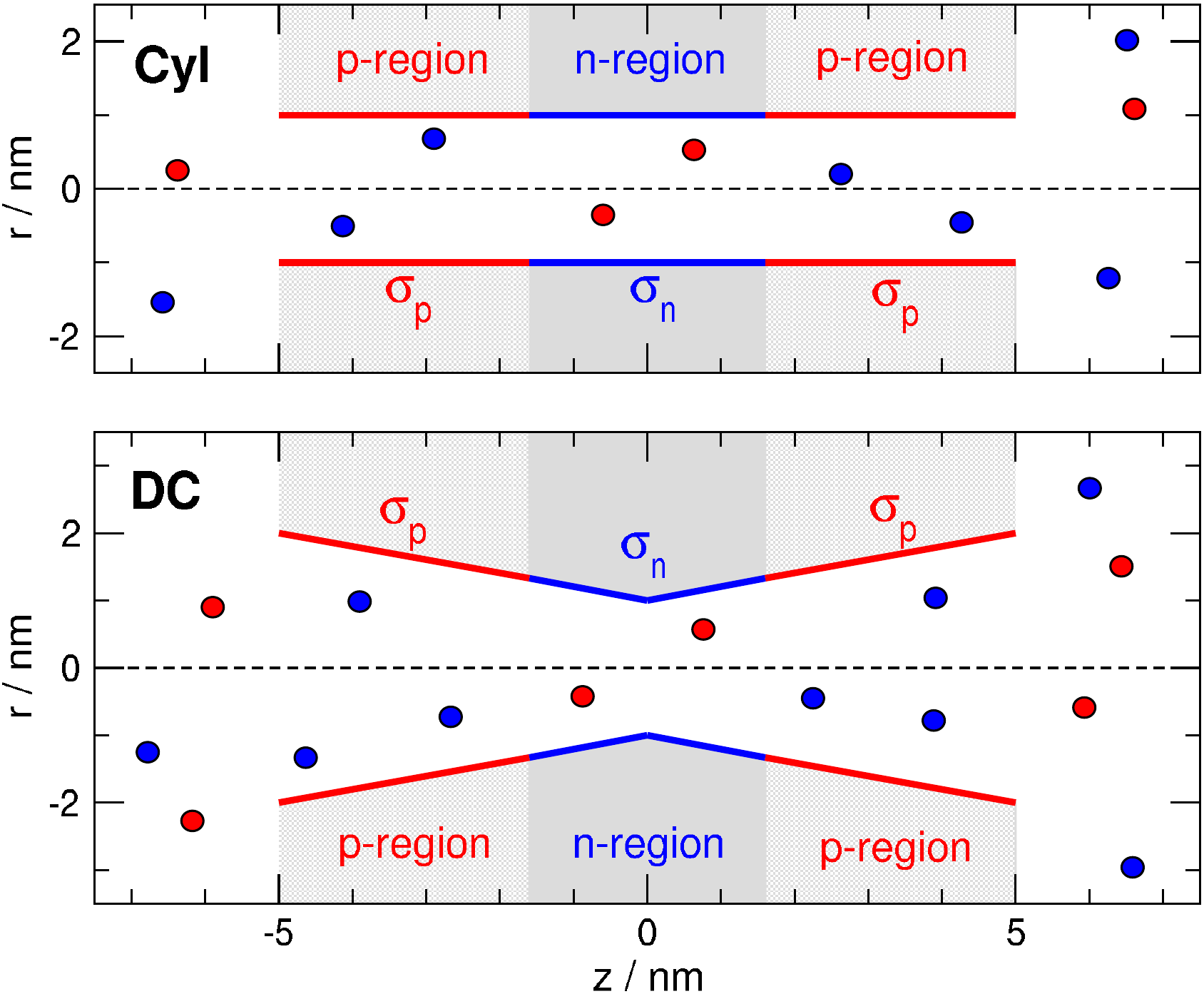}}}
\end{center}
\caption{Geometries for the cylindrical (Cyl, top) and double conical (DC, bottom) nanopores. The figure shows the situation in the OFF state ($\mathrm{pH}=7$), when $\sigma_{\mathrm{p}}=1$ $e$/nm$^{2}$ and $\sigma_{\mathrm{n}}=-1$ $e$/nm$^{2}$.  Red and blue colors indicate negative (anions) and positive (cations) charges, respectively, from now on.
The full simulation domain is larger than that shown in the figure.}
\label{fig1}
\end{figure}

We applied two pore geometries.
The cylindrical (Cyl) pore has radius $R_{\mathrm{pore}}=1$ nm. (Fig.\ \ref{fig1}, top panel).
The double conical (DC) pore has a radius $R_{\mathrm{pore}}(z=0)=1$ nm in the center, while it has $R_{\mathrm{pore}}(z=\pm5)=2$ nm at the entrances (Fig.\ \ref{fig1}, bottom panel).
The length of the pores is $H_{\mathrm{pore}}=10$ nm that is small compared to experimental values for polymeric nanopores, but our model calculation's goal is to study the effect of pH-dependent surface charge pattern.
This effect is qualitatively the same for longer pores as was shown in our previous work \cite{madai_pccp_2018}, where pore length dependence was considered.

The membrane and the pore is confined by hard walls with which the overlap of ions is forbidden.
A symmetric charge pattern is created on the wall of the nanopore with regions on the two ends of the pore carrying $\sigma_{\mathrm{p}}$ surface charges (p-regions), while there is a central region of 3.2 nm width carrying charge $\sigma_{\mathrm{n}}$ (n-region).
The values of these surface charges can be regulated with pH as detailed in the next subsection.

This surface charge is represented by collections of fractional point charges placed on a rectangular grid.
The width of the grid elements is 0.2 nm, while the magnitudes of the point charges are determined by the values $\sigma_{\mathrm{p}}$ and $\sigma_{\mathrm{n}}$.

\subsection{pH-dependence of surface charges}

Surface charge is adjusted through protonation and deprotonation of the anchored functional groups.
Negative charge comes from deprotonated acidic groups (A$^{-}$, carboxyl or sulphate \cite{lepoitevin_rscadv_2016}), while positive charge comes from protonated basic groups (BH$^{+}$, amino/imino or pyridinium groups \cite{yameen_chemcomm_2010}).
Surface charges corresponding to A$^{-}$ and BH$^{+}$ groups can be expressed as
\begin{align}
\sigma_{\mathrm{n}} &= \Phi^{\mathrm{A}^-} \sigma_{\mathrm{n}}^{\mathrm{max}}   \\ \nonumber
\sigma_{\mathrm{p}}&  = \Phi^{\mathrm{BH}^+} \sigma_{\mathrm{p}}^{\mathrm{max}}  , 
\end{align}
where $\Phi^{\mathrm{A}^{-}}$ and $\Phi^{\mathrm{BH}^{+}}$ are the mole fractions of the corresponding charged groups on the surface of the nanopore at a given pH, and $\sigma_{\mathrm{n}}^{\mathrm{max}} $ and $\sigma_{\mathrm{p}}^{\mathrm{max}}$ are the maximal values of surface charges depending on the densities of the functional groups on the surface.
In this study, we use the values $\sigma_{\mathrm{n}}^{\mathrm{max}}=-1 $ $e$/nm$^{2}$ and $\sigma_{\mathrm{p}}^{\mathrm{max}}=1$ $e$/nm$^{2}$ that are realistic in experiments \cite{kalman_am_2008}.

The mole fractions, on the other hand, depend on pH and the corresponding pK$_{\mathrm{a}}$ (logarithmic acid dissociation constant) values of the functional groups via
\begin{align}
\label{sign}
\Phi^{\mathrm{A}^-} & = \dfrac{1}{1+10^{-(\mathrm{pH}-\mathrm{pK}_{\mathrm{a}}^{\mathrm{A}^{-}})}}  \\ \nonumber
\Phi^{\mathrm{BH}^+} & = \dfrac{1}{1+10^{(\mathrm{pH}-\mathrm{pK}_{\mathrm{a}}^{\mathrm{BH}^{+}})}}
,
\end{align}
where $\mathrm{K}_{\mathrm{a}}^{\mathrm{A}^{-}}$ and $\mathrm{K}_{\mathrm{a}}^{\mathrm{BH}^{+}}$ are the equilibrium constants of the dissociation reactions  $\mathrm{AH}\rightleftharpoons \mathrm{A}^{-} + \mathrm{H}^{+}$ and $\mathrm{BH}^{+} \rightleftharpoons \mathrm{B} + \mathrm{H}^{+}$.
For the pH regions, where 99\% of the groups are protonated/deprotonated we use the following approximations: 
\begin{itemize}
\item $\Phi^{\mathrm{A}^-}=0$ for $\mathrm{pH}\leq \mathrm{pK}_{\mathrm{a}}^{\mathrm{A}^{-}}-2$, 
\item $\Phi^{\mathrm{A}^-}=1$ for $\mathrm{pH}\geq \mathrm{pK}_{\mathrm{a}}^{\mathrm{A}^{-}}+2$, 
\item $\Phi^{\mathrm{BH}^+}=0$ for $\mathrm{pH}\leq \mathrm{pK}_{\mathrm{a}}^{\mathrm{BH}^{+}}+2$, 
\item $\Phi^{\mathrm{BH}^+}=1$ for $\mathrm{pH}\geq \mathrm{pK}_{\mathrm{a}}^{\mathrm{BH}^{+}}-2$.                                                                                                                                                                                                                                                                                                                                                                                   \end{itemize}
For the dissociation constants,  we used the values pK$_{\mathrm{a}}^{\mathrm{A}^{-}}=5$ \cite{hou_advmat_2010} and pK$_{\mathrm{a}}^{\mathrm{BH}^{+}}=9$ \cite{Ali_ACSNano_2009} characteristic of -COOH and -NH$_{3}^{+}$ groups \cite{tagliazucchi_book_2017}.
The curves for the pH-dependence of the surface charges using these parameters are shown in Fig.\ \ref{fig2}A.
This treatment is similar to that used by others in modeling studies \cite{yeh_analchem_2013,xue_jpcc_2014,hsu_analchem_2017}.

We introduce special cases for combinations of $\sigma_{\mathrm{n}}$ and $\sigma_{\mathrm{p}}$.
We denote 
the case $\sigma_{\mathrm{p}}=1$ $e/\mathrm{nm}^{2}$ and $\sigma_{\mathrm{n}}=0$ $e/\mathrm{nm}^{2}$ with ``$+0+$'',
the case $\sigma_{\mathrm{p}}=1$ $e/\mathrm{nm}^{2}$ and $\sigma_{\mathrm{n}}=-1$ $e/\mathrm{nm}^{2}$ with ``$+-+$'', and 
the case $\sigma_{\mathrm{p}}=0$ $e/\mathrm{nm}^{2}$ and $\sigma_{\mathrm{n}}=-1$ $e/\mathrm{nm}^{2}$ with ``$0-0$'' as depicted in Fig.\ \ref{fig2}A.

\subsection{Reduced model of electrolyte and ion transport}

The interactions and transport of ions are described with a reduced model.
In this approach, the degrees of freedom of water molecules are coarse-grained into response functions that characterize the effects of water molecules on ions.
We consider two types of effects.

First, water molecules screen the charges of ions. 
This is an ``energetic'' effect in nature because it determines the energy of the system through the pair-potential acting between the charged hard spheres with which we model the ions:
\begin{equation}
 u_{ij}(r)
=\left\{
        \begin{array}{ll}
    \infty & \; \mbox{for} \; \;  r<R_{i}+R_{j}\\
        \dfrac{q_{i}q_{j}}{4\pi \epsilon_{0}\epsilon r} & \; \mbox{for} \; \; r \geq R_{i}+R_{j}  ,
        \end{array}
        \right. 
\label{eq:pm}
\end{equation} 
where $q_{i}=\pm e$ and $R_{i}$ are the charge and radius of ionic species $i$, respectively, $\epsilon_{0}$ is the permittivity of vacuum, and $r$ is the distance between the ions.
Screening is taken into account by a dielectric constant, $\epsilon$, in the denominator of the Coulomb potential.

Second, water molecules hinder the diffusion of ions via friction \cite{einstein1956,bockris-reddy,robinson-stokes}.
This is a ``dynamic'' effect in nature because it influences the flux of ions through the Nernst-Planck (NP) transport equation:
\begin{equation}
 -kT\mathbf{j}_{i}(\mathbf{r}) = D_{i}(\mathbf{r})c_{i}(\mathbf{r})\nabla \mu_{i}(\mathbf{r}),
 \label{eq:np}
\end{equation} 
where $k$ is Boltzmann's constant, $T$ is temperature, $\mathbf{j}_{i}(\mathbf{r})$ is the particle flux density of ionic species $i$, $c_{i}(\mathbf{r})$ is the concentration, and $\mu_{i}(\mathbf{r})$ is the electrochemical potential profile.
Due to rotational symmetry, the functions in the NP equation have the two cylindrical variables $z$ and $r$: $c_{i}(z,r)$, $ \mu_{i}(z,r)$, and $\mathbf{j}_{i}(z,r)$.

The response function is the diffusion coefficient profile, $D_{i}(\mathbf{r})$, in this case.
It is a parameter that can be adjusted either to experiments (as in the case of the Ryanodine receptor calcium channel \cite{boda-arcc-2014,fertig-hjic-45-73-2017}) or to results of molecular dynamics simulations (as in the case of our study for bipolar nanopores \cite{hato-pccp-19-17816-2017}).
We assume that it changes only along the $z$-dimension: $D_{i}(z)$.

We want the two ions to be identical apart from their charges in our model calculations, because we want to focus on interactions of ions with surface charges and applied field. 
At this point, we are not interested in the effects of asymmetries in the behaviors of ionic species.
Therefore, the bulk diffusion constant of both ion species is $1.334\cdot10^{-9}$ m$^{2}$/s, while the value inside the pore is ten times smaller \cite{matejczyk-jcp-146-124125-2017,madai-jcp-147-244702-2017}, a choice that does not qualitatively affect our conclusions.
Also, we use the same radius for both ions, $R_{i}=0.15$ nm.

To solve the NP equation, we need a closure between the concentration profile, $c_{i}(\mathbf{r})$, and the electrochemical potential profile, $\mu_{i}(\mathbf{r})$.
This closure is generally provided  by statistical mechanics. 
In this work, we use the Local Equilibrium Monte Carlo (LEMC), which is a particle simulation method described in the next subsection.
Once the relation between $c_{i}(\mathbf{r})$ and  $\mu_{i}(\mathbf{r})$ is available, a self-consistent solution is obtained iteratively in which the conservation of mass, namely, the continuity equation, $\nabla \cdot \mathbf{j}_{i}(\mathbf{r}) = 0$, is satisfied \cite{boda-jctc-8-824-2012,boda-jml-189-100-2014,boda-arcc-2014,fertig-hjic-45-73-2017}.

\subsection{Local Equilibrium Monte Carlo}
\label{subseq:lemc}

LEMC is an adaptation of the Grand Canonical Monte Carlo (GCMC) technique to a non-equilibrium situation \cite{boda-jctc-8-824-2012,boda-jml-189-100-2014,boda-arcc-2014,fertig-hjic-45-73-2017}.
The independent state function of the LEMC simulation is the chemical potential profile, $\mu_{i}(\mathbf{r})$, while the output variable is the concentration profile, $c_{i}(\mathbf{r})$.
Chemical potential is constant in space in global equilibrium for which GCMC simulations were originally designed.
Out of equilibrium, however, $\mu_{i}(\mathbf{r})$ is a space-dependent quantity.

The transition from global equilibrium to non-equilibrium is possible by assuming local equilibrium (LE).
We divide the solution domain into small volume elements, $\mathcal{B}^{\alpha}$, and assume that the chemical potential is constant in this volume, $\mu_{i}^{\alpha}$. 
This function tunes the probability that ions of species $i$ occupy this volume.
LEMC applies ion insertion/deletion steps that are very similar to those used in global-equilibrium GCMC \cite{boda-jctc-8-824-2012} with the differences that (1) the electrochemical potential value, $\mu_{i}^{\alpha}$, assigned to the volume element in which the insertion/deletion happens, $\mathcal{B}^{\alpha},$ is used in the acceptance probability, (2) the volume of the volume element, $V^{\alpha}$, is used in the acceptance probability, and (3) the energy change associated with the insertion/deletion, $\Delta U$, contains all the interactions from the whole simulation cell, not only from volume element $\mathcal{B}^{\alpha}$.

The advantage of the technique in which LEMC is coupled to NP (coined as NP+LEMC) is that it correctly computes volume exclusion and electrostatic correlations between ions, so it is beyond the mean-field level of the PNP theory. 
Its advantage compared to the Brownian Dynamics method \cite{chung-bj-77-2517-1999,im_bj_2000,berti-jctc-10-2911-2014} is that sampling of ions passing the pore is not necessary: current is computed from the NP equation.
Sampling of passing ions can be poor especially when these events are rare due to the small current associated with the depletion zones of ions.
The transistors studied here belong to this category because their behavior is governed by depletion zones. 

Our NP+LEMC method has been successfully applied for membranes \cite{boda-jctc-8-824-2012,hato-jcp-137-054109-2012}, ion channels \cite{boda-jml-189-100-2014,boda-arcc-2014,fertig-hjic-45-73-2017,hato-cmp-19-13802-2016} and nanopores \cite{hato-pccp-19-17816-2017,matejczyk-jcp-146-124125-2017,madai-jcp-147-244702-2017,madai_pccp_2018}.
A comparison with molecular dynamics simulations revealed that the implicit water model can capture the device properties of nanopores with variable charge patterns \cite{hato-pccp-19-17816-2017}, while comparisons with PNP showed that PNP can reproduce device function qualitatively despite quantitative inaccuracies for 1:1 electrolytes \cite{matejczyk-jcp-146-124125-2017,madai_pccp_2018}.

\section{Results and Discussion}

\begin{figure}[t]
\begin{center}
\rotatebox{0}{\scalebox{0.7}{\includegraphics*{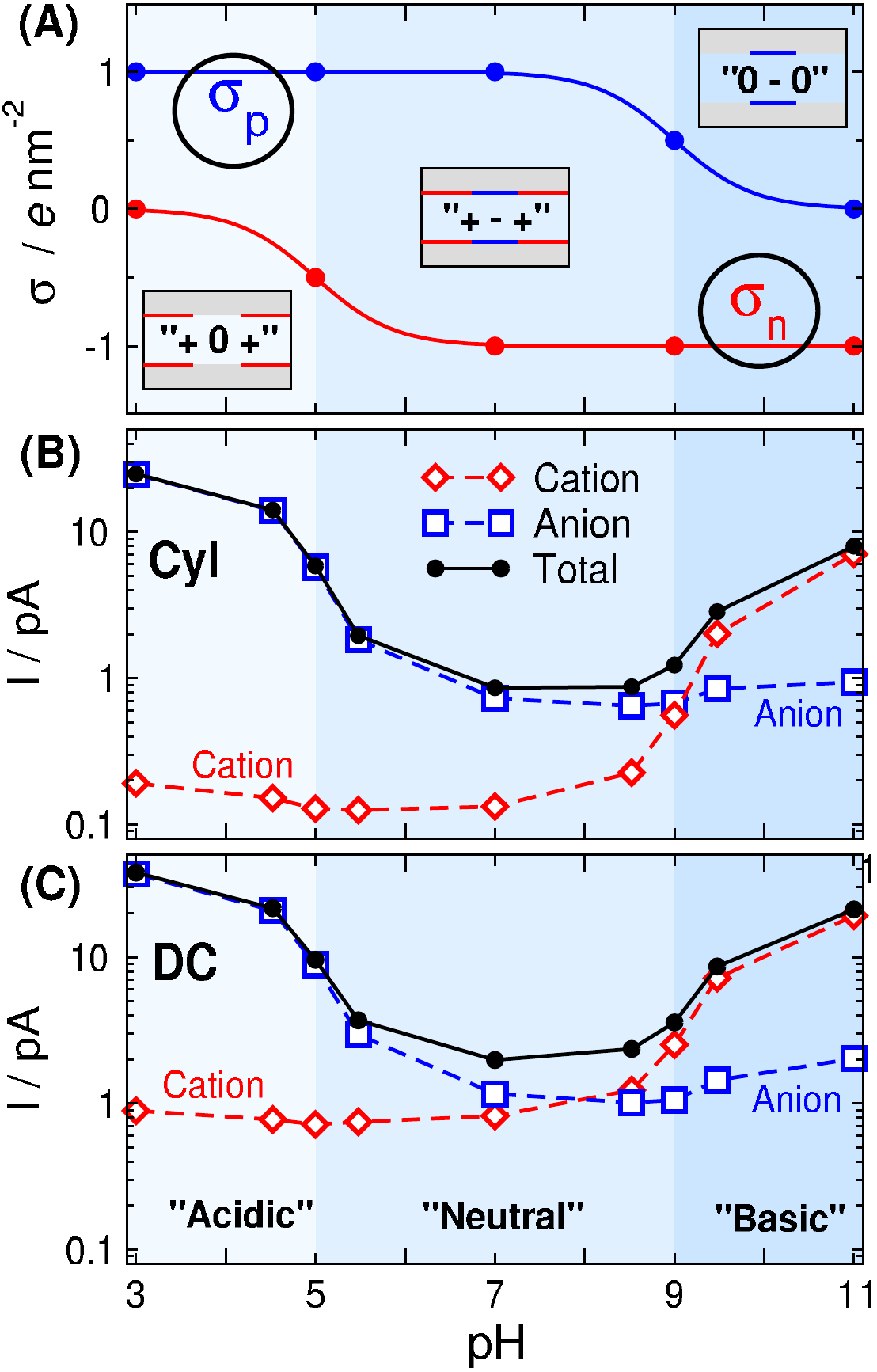}}}
\end{center}
\caption{(A) pH-dependence of $\sigma_{\mathrm{n}}$ and $\sigma_{\mathrm{p}}$ using parameters $\sigma_{\mathrm{n}}^{\mathrm{max}}=-1 $ $e$/nm$^{2}$, $\sigma_{\mathrm{p}}^{\mathrm{max}}=1$ $e$/nm$^{2}$,  pK$_{\mathrm{a}}^{\mathrm{A}^{-}}=5$, and pK$_{\mathrm{a}}^{\mathrm{BH}^{+}}=9$.
The region below pK$_{\mathrm{a}}^{\mathrm{A}^{-}}=5$ (light blue) is indicated as ``acidic'', while the region above pK$_{\mathrm{a}}^{\mathrm{BH}^{+}}=9$ (dark blue) is indicated as ``basic''.
In between, indicated as ``neutral'', both surface charges are below $\pm 0.5$ $e$/nm$^{2}$.
These regions shown with different colors are characterized with the terms in apostrophes (``acidic'', ``basic'', ``neutral'') with meanings  possibly different from the usual chemical ones.
(B) and (C) Currents of cation, anion, and theirs sum as functions of pH for the cylindrical (B) and double conical (C) geometry. Note the logarithmic scale of the ordinate.}
\label{fig2}
\end{figure}

We expect from our device that its response to pH will be characteristically different at the different pH regions depicted in Fig.\ \ref{fig2}A.
Indeed, as Figs.\ \ref{fig2}B and C show, the current is considerable in the ``acidic'' and ``basic'' regions (ON state of the device), while it is very small in the ``neutral'' region (OFF state of the device).
We emphasize that the terms between the apostrophes do not correspond rigidly to the usual chemical terms; they refer to the regions indicated by different colors in Fig.\ \ref{fig2}.

In the ``acidic'' (light blue) region, the positive charge of the p-regions ($\sigma_{\mathrm{p}}=1$ $e$/nm$^{2}$) attracts anions into the pore, while the moderate negative charge of the n-region ($-0.5\leq\sigma_{\mathrm{n}}\leq0$ $e$/nm$^{2}$) cannot exclude them.
Therefore, anions carry current in this case.
Meanwhile, cations are effectively excluded by the $\sigma_{\mathrm{p}}=1$ $e$/nm$^{2}$  charge in the p-regions, their current, therefore, is small (anion selectivity).

\begin{figure}[t!]
\begin{center}
\rotatebox{0}{\scalebox{0.65}{\includegraphics*{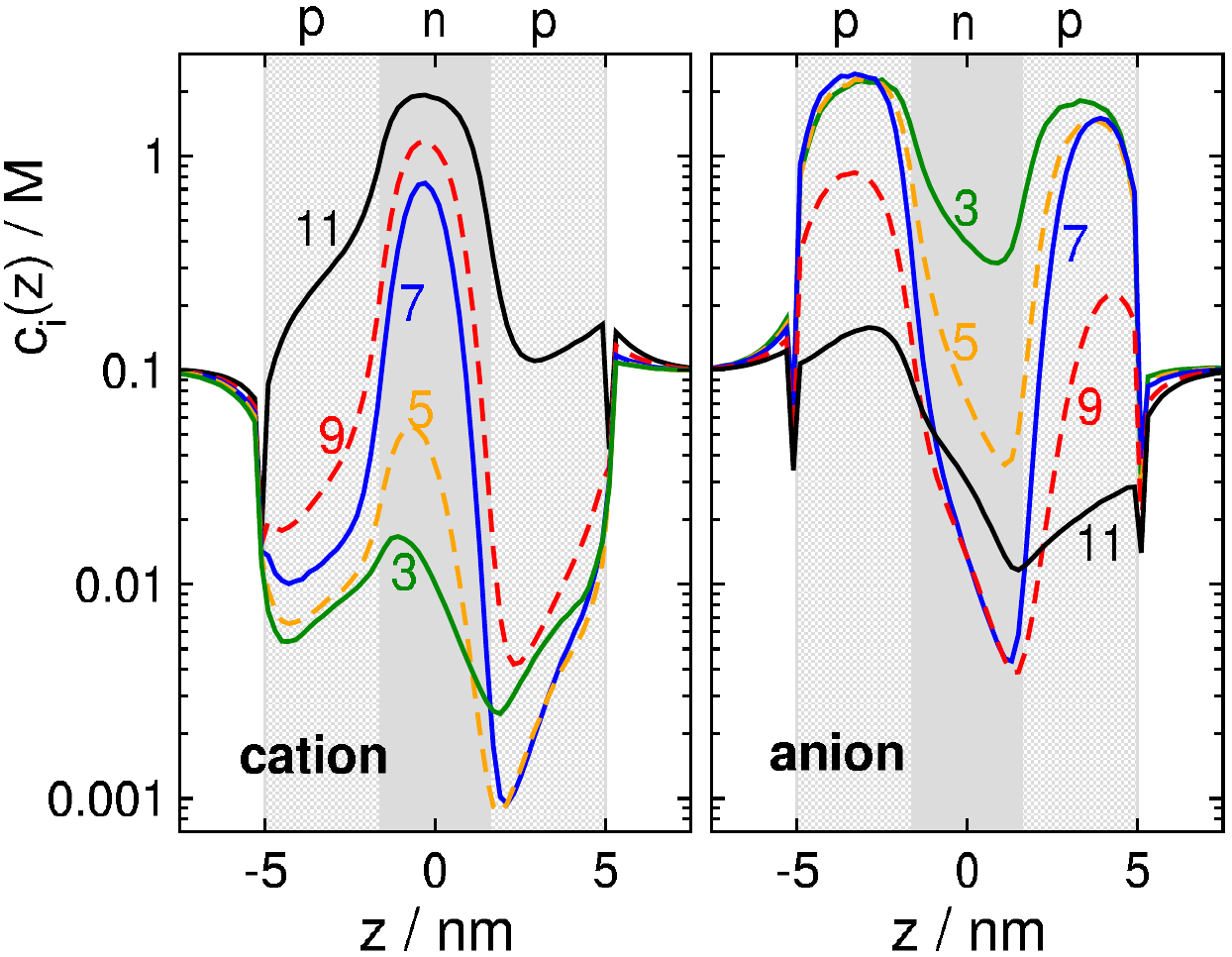}}}
\end{center}
\caption{Concentration profiles of cations (left panel) and anions (right panel) for different pH values from 3 to 11 (see numbers near curves) for the Cyl geometry.}
\label{fig3}
\end{figure}

This is well explained by the concentration profiles in Fig.\ \ref{fig3}.
At $\mathrm{pH}=3$ (green lines), cations  (left panel) have depletion zones in the p-regions where $\sigma_{\mathrm{p}}=1$ $e$/nm$^{2}$, while  anions (right panel) accumulate in these zones and even in the n-region, where $\sigma_{\mathrm{n}}=0$ $e$/nm$^{2}$.

Increasing pH to 5,  $\sigma_{\mathrm{p}}=1$ $e$/nm$^{2}$ as for $\mathrm{pH}=3$ so cations are still excluded from the p-regions (orange curves).
Anions, on the other hand, are present in a smaller concentration in the central n-region due to its increased charge ($\sigma_{\mathrm{n}}=-0.5$ $e$/nm$^{2}$).
This results in a smaller anion current.

At $\mathrm{pH}=7$, both ions have depletion zones in the regions that repel them: cations in the p-zones, while anions in the n-zone (blue curves).
Due to the depletion zones, neither of the ionic species carries current, at least, not much. 
A depletion zone of an ionic species anywhere along the ionic pathway inside the pore cuts the current of that species because the consecutive segments of the pore behave as resistors connected in series.
If any of the resistors has a large resistance due to the absence of charge carriers there (the depletion zone), the resistance of the whole circuit will be large.

The behavior on the ``basic'' side (above $\mathrm{pH}=9$) is similar, but opposite compared to the ``acidic'' side.
At $\mathrm{pH}=11$ (black lines), anions have depletion zones in the n-region where $\sigma_{\mathrm{n}}=-1$ $e$/nm$^{2}$  (right panel), while  cations accumulate in this zone and even in the p-region, where $\sigma_{\mathrm{p}}=0$ $e$/nm$^{2}$ (left panel).
On this side of the pH regime, therefore, the main charge carriers are the cations (cation selectivity).

Decreasing pH to 9,  $\sigma_{\mathrm{n}}=-1$ $e$/nm$^{2}$ as for $\mathrm{pH}=11$ so anions are still excluded from the n-region (red curves).
Cations, on the other hand, are present in a smaller concentration in the p-regions due to the increased charge ($\sigma_{\mathrm{p}}=0.5$ $e$/nm$^{2}$) of the p-regions.
This results in a smaller cation current.

Although the charge pattern is symmetric, the concentration profiles in Fig.\ \ref{fig3} are asymmetric.
This is due to the external electric field imposed on the system (200 mV on the right hand side, ground on the left).

Fig.\ \ref{fig2}C shows the current data for the DC geometry.
The basic behavior is the same as in the Cyl case.
There are, however, differences due to different geometries.

While the pore radius is the same at the center ($z=0$) in both geometries, the pore gets wider approaching the entrances at both sides in the DC geometry.
The effect of this can be shown by plotting line-density profiles, $n_{i}(z)$, instead of concentration profiles, $c_{i}(z)$.
While $c_{i}(z)$ is the average of $c_{i}(z,r)$ over a cross section, $n_{i}(z)$ is the integral of $c_{i}(z,r)$ over the cross section, therefore, it characterizes the average number of ions at a given point along the $z$-axis:
\begin{equation}
n_{i}(z) = 2\pi \int_{A(z)}c_{i}(z,r)r \mathrm{d}r = c_{i}(z) A(z),
\label{eq:n}
\end{equation}
where $A(z)$ is the cross section of the pore at $z$.
Therefore, line-density shows that there are more ions in the wider regions of the pore in the DC geometry.

The current is the cross sectional integral of the flux: 
\begin{equation}
 I_{i}=q_{i}2\pi \int_{A(z)} \mathbf{j}_{i}(z,r)\cdot \mathbf{k}\,r\, \mathrm{d}r ,
 \label{eq:I}
\end{equation} 
where $\mathbf{k}$ is the unit vector in the $z$-dimension.
Replacing the NP equation (Eq.\ \ref{eq:np}) for $\mathbf{j}_{i}(z,r)$ and assuming that the $r$-dependence of $\mu_{i}(z,r)$ is weak (this is true up to a good approximation; results not shown), we obtain that
\begin{equation}
I_{i}=-q_{i} \dfrac{D_{i}(z)}{kT}\left[ 2\pi \int_{A(z)} c_{i}(z,r)\, r \,\mathrm{d}r \right] \dfrac{\mathrm{d}\mu_{i}(z)}{\mathrm{d}z}  ,
\end{equation} 
and, using Eq.\ \ref{eq:n} for the expression in the square bracket,
\begin{equation}
 I_{i} = -q_{i} \dfrac{D_{i}(z)}{kT} n_{i}(z) \dfrac{\mathrm{d}\mu_{i}(z)}{\mathrm{d}z}  .
 \label{eq:i-n}
\end{equation} 
The value of the current is the same for all $z$ values inside the pore according to conservation of mass.
This result shows that, in this approximation, the current depends on $n_{i}$ and the slope of $\mu_{i}$.

\begin{figure}[t!]
\begin{center}
\rotatebox{0}{\scalebox{0.65}{\includegraphics*{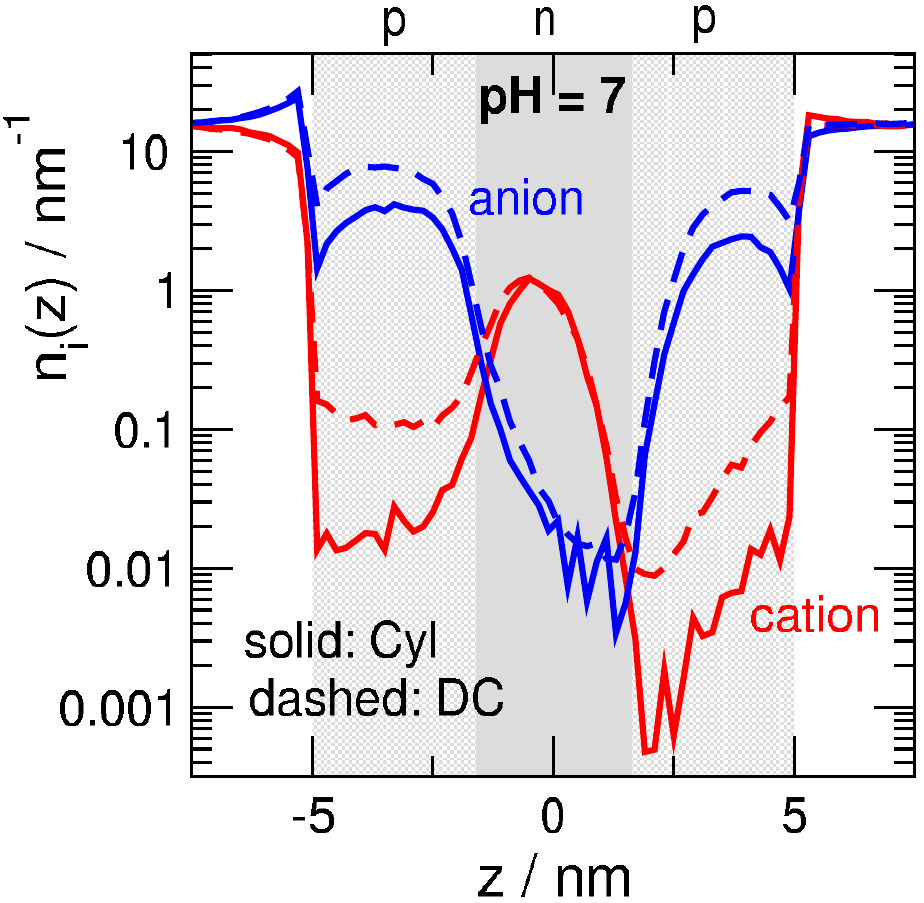}}}
\end{center}
\caption{Line-density profiles (see Eq.\ \ref{eq:n}) of cations (red) and anions (blue) in the OFF state ($\mathrm{pH}=7$). Solid and dashed curves represent results for the Cyl and DC geometries, respectively.}
\label{fig4}
\end{figure}

There are two basic observations from Fig.\ \ref{fig2}B and C regarding differences between the two geometries.
First, cation currents are smaller in the Cyl geometry for $\mathrm{pH}\leq7$. 
This is because the pore is narrower at the two entrances in the Cyl case, so the $\sigma_{\mathrm{p}}=1$ $e$/nm$^{2}$ surface charge repulses cations more efficiently.
To show this, we plot the line-density profiles for $\mathrm{pH}=7$ in Fig.\ \ref{fig4}.
The cation $n_{i}(z)$ profiles are much deeper in the p-regions in the Cyl geometry (see solid red lines vs.\ dashed red lines), so, due to Eq.\ \ref{eq:i-n}, cation currents are smaller.
Anion currents are less influenced by geometry in this pH regime.

\begin{figure}[t!]
\begin{center}
\rotatebox{0}{\scalebox{0.6}{\includegraphics*{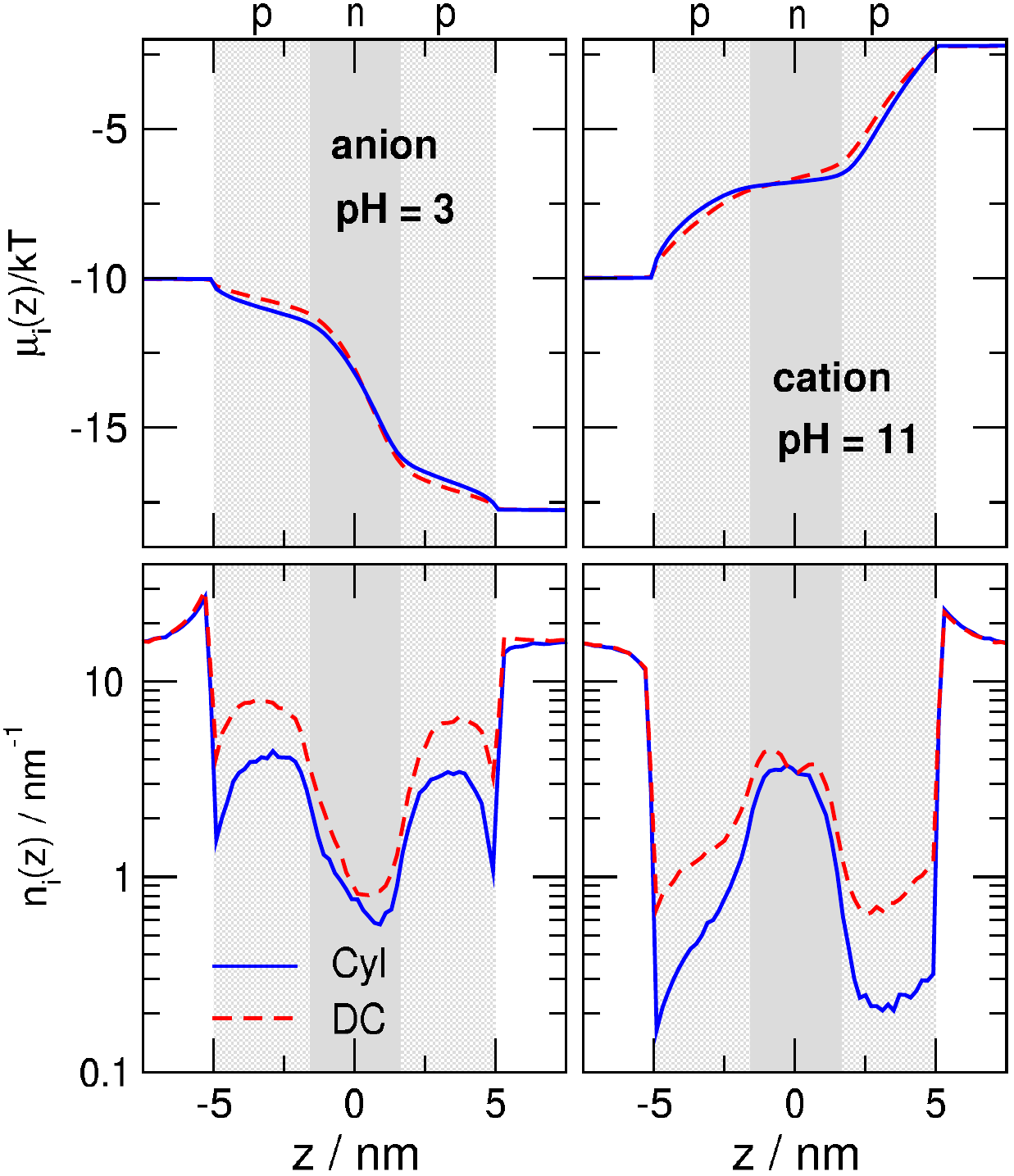}}}
\end{center}
\caption{Electrochemical potential (top) and line-density profiles (botom) of charge carrying ions in the open states for $\mathrm{pH}=3$ (left) and $\mathrm{pH}=11$ (right). In the ``acidic'' ($\mathrm{pH}=3$) case the anions, while in the ``basic''  ($\mathrm{pH}=11$) case the cations are the charge carriers. Solid blue and dashed red curves represent results for the Cyl and DC geometries, respectively.}
\label{fig5}
\end{figure}

The second observation is that currents of the charge-carrying species are larger in the DC geometry.
This phenomenon cannot be explained on the basis of the $n_{i}(z)$ profiles alone.
In the center at the bottleneck ($z=0$), the line densities are quite similar in the DC and Cyl geometries, although they are somewhat larger in the DC geometry (bottom row of Fig.\ \ref{fig5}).
However, if we also take the driving force, $\mathrm{d}\mu_{i}(z)/\mathrm{d}z$, into account, an additional factor appears that increases the current in the DC geometry.
Indeed, the top row of Fig.\ \ref{fig5} shows that the $\mu_{i}(z)$ curve is steeper in the DC geometry at $z=0$.
Therefore, the current is larger also due to the larger driving force at $z=0$.
In the p-regions, the $n_{i}(z) \, \mathrm{d}\mu_{i}(z)/\mathrm{d}z$ product remains the same, because $n_{i}(z)$ is larger, while $ \mathrm{d}\mu_{i}(z)/\mathrm{d}z$ is smaller in the DC geometry.

\section{Summary}

In this work, we considered a model device that makes it possible to control ionic current through a nanopore by simply changing the pH of the electrolyte.
pH sets the surface charges through the protonation/deprotonation states of the functional groups with which the pore surface is functionalized.

Current is controlled  by the pH-dependent surface charge pattern.
If both positive and negative charges are present on the pore wall, both ions have depletion zones.
In this case, currents of both ions are limited by these depletion zones.
This case corresponds to the OFF state.
In the ON state, the depletion zone of one of the ionic species is absent.
That species carries the current in the ON state.
Since the other species is still excluded, the pore is ion selective in the ON state.
This shows that our nanopore model is pH-gated because it is either open or closed depending on the value of pH.

Our work confirms that nanopores can be used for bridging chemistry and electronic circuitries by changing the surface charge pattern with chemical treatment and by the presence of certain ions (H$^{+}$, in this case).
Change in the surface charge, in turn, changes the conductance properties of the pore.
This principle makes it possible to use nanopores as sensors.

\section*{Acknowledgements}
\label{sec:ack}

We gratefully acknowledge  the financial support of the National Research, Development and Innovation Office -- NKFIH K124353. 
We acknowledge the financial support of Sz\'echenyi 2020 under the EFOP-3.6.1-16-2016-00015.

%

\end{document}